\documentclass{aa}  
\usepackage{graphicx}
\usepackage{txfonts}
\usepackage{natbib,twoopt}
\usepackage[breaklinks=true]{hyperref} 
\bibpunct{(}{)}{;}{a}{}{,}             
\makeatletter
  \newcommandtwoopt{\citeads}[3][][]{\href{http://adsabs.harvard.edu/abs/#3}%
    {\def\hyper@linkstart##1##2{}%
     \let\hyper@linkend\@empty\citealp[#1][#2]{#3}}}
  \newcommandtwoopt{\citepads}[3][][]{\href{http://adsabs.harvard.edu/abs/#3}%
    {\def\hyper@linkstart##1##2{}%
     \let\hyper@linkend\@empty\citep[#1][#2]{#3}}}
  \newcommandtwoopt{\citetads}[3][][]{\href{http://adsabs.harvard.edu/abs/#3}%
    {\def\hyper@linkstart##1##2{}%
     \let\hyper@linkend\@empty\citet[#1][#2]{#3}}}
  \newcommandtwoopt{\citeyearads}[3][][]%
    {\href{http://adsabs.harvard.edu/abs/#3}
    {\def\hyper@linkstart##1##2{}%
     \let\hyper@linkend\@empty\citeyear[#1][#2]{#3}}}
\makeatother
\def\la{\mathrel{\hbox{\rlap{\hbox{\lower4pt\hbox{$\sim$}}}\hbox{$<$}}}}
\def\ga{\mathrel{\hbox{\rlap{\hbox{\lower4pt\hbox{$\sim$}}}\hbox{$>$}}}}

\def\deg      {{\ifmmode^\circ\else$^\circ$\fi} } 
\def\arcmin   {{\ifmmode {'}\else$'$\fi}} 
\def\arcsec   {{\ifmmode{''}\else$''$\fi}} 

\begin{document}

\title{The stability of radially anisotropic rotating stellar systems with a central density cusp}
\authorrunning{Di Cintio \& Varri}
\titlerunning{ROI in cuspy rotating stellar systems}
 \author{Pierfrancesco Di Cintio
          \inst{1,2,3}
          \and
         Anna Lisa Varri
          \inst{4,5}
          }
   \institute{Consiglio Nazionale delle Ricerche, Istituto dei Sistemi Complessi, via Madonna del Piano 17, 50022 Sesto Fiorentino (FI), Italy\\
              \email{pierfrancesco.dicintio@cnr.it}
         \and
             INFN -- Sezione di Firenze, via G. Sansone 1, 50022 Sesto Fiorentino (FI), Italy
         \and
             INAF -- Osservatorio Astrofisico di Arcetri, Largo E. Fermi 5, 50125 Firenze, Italy   
         \and
               School of Mathematics and Maxwell Institute for Mathematical Sciences, University of Edinburgh, King's Buildings,
               Edinburgh EH9 3FD, UK
           \and
           Institute for Astronomy, University of Edinburgh, Royal Observatory, Blackford Hill, Edinburgh EH9 3HJ, UK
            \email{anna.varri@ed.ac.uk}
             }
   \date{Received ...; accepted ...}

 
\abstract
   {}
   {We investigate the interplay between radial velocity anisotropy and internal rotation in self-gravitating systems characterised by initial spherical symmetry and a moderate density cusp. We study the stability properties of such configurations to disentangle the impact of anisotropy, rotation, and density structure on the formation of triaxial stellar systems.} 
   {We perform a set of collisionless $N$-body simulations starting from initial equilibria with a phase space distribution function of the class introduced by Osipkov and Merritt, modified to impose a net global angular momentum with the Lynden-Bell daemon protocol. We analyse the growth of the density and phase space distribution modes, as well as the axial ratios and triaxiality index of the final configurations.}
   {We find that internal rotation has a mitigating effect on the strength of the radial orbit instability in radially anisotropic models with \mbox{Fridman-Polyachenko-Shukhman} anisotropy index close to the consistency limit, while it slightly enhances the onset of the instability in marginally stable models. From the analysis of the growth rate of the density modes, we also find that an inner bar, which for maximally rotating initial conditions declines after its initial growth, is formed even in systems with initial profiles with a central density cusp.}
   {}

\keywords{galaxies: formation -- galaxies: evolution -- gravitation -- methods: analytical -- methods: numerical }
\maketitle
\nolinenumbers
\section{Introduction}\label{intro}
Spherical systems characterised by a significant amount of kinetic energy stored in radial motion (i.e., low angular momentum orbits) are known to be violently unstable under the process of radial orbit instability (hereafter ROI, see, e.g., \citealt{1981SvA....25..533P,1987MNRAS.224.1043P,1992SvA....36..482P,2017MNRAS.470.2190P}). The nature of ROI as a collective phenomenon (\citealt{2011TTSP...40..425M}) such as, for example, \cite{1967MNRAS.136..101L} ``violent relaxation'' (hereafter VR), or, by contrast, as a process associated with the amplification of a finite set of unstable modes via local discreteness effects (\citealt{1991ApJ...368...66W,1994ApJ...434...94B}), is still unclear. Notwithstanding such an ambiguity, a qualitative interpretation of ROI in terms of an accumulation of particles on nearly radial orbits reverting their precession frequency onto a pre-existing aspherical perturbation is usually accepted (\citealt{2015MNRAS.451..601P}), and is qualitatively akin to the \cite{1979MNRAS.187..101L} mechanism of bar formation in disk galaxies (\citealt{1991MNRAS.248..494S,2020MNRAS.498.3368P}).\\
\indent Although extensively studied for anisotropic equilibrium models (see, e.g., \citealt{2011MNRAS.416.1836P} and references therein), ROI is also at play during the evolution of initially far-from-equilibrium systems (\citealt{1994ApJ...424..106H,2005A&A...433...57T}) undergoing VR, such as the early phases of the collapse of initially cold systems. Therefore, ROI is often invoked as the mechanism that gives rise to triaxiality in self-gravitating systems that undergo dissipationless collapse from nearly spherical and cold initial conditions (\citealt{1988CeMec..41....3A,1999AJ....118.2158B,2005ApJ...634..775B,2008ApJ...685..739B}), where an initially minor departure from spherical symmetry grows by accreting particles on low angular momentum orbits (\citealt{2011TTSP...40..425M}). This notion is empirically confirmed by numerical experiments (see, among others, \citealt{1985MNRAS.217..787M,1986ApJ...300..112B,1990MNRAS.242..576A,1996MNRAS.280..700P,1997ApJ...490..136M,2002MNRAS.332..901N,2009ApJ...704..372B,2015MNRAS.447...97G,2020MNRAS.494.1027D}).\\
\indent It is well established that, for a given functional form of the density profile, ROI is stronger in models bearing a larger degree of radial anisotropy, which is usually quantified (see, e.g., \citealt{2008gady.book.....B}, as well as \citealt{1984sv...bookR....F} and Sect.~\ref{mod1} below) by the so-called \mbox{Fridman-Polyachenko-Shukhman} anisotropy index $\xi$ that is proportional to the ratio of the radial and tangential components of the kinetic energy tensor. Whereas ROI is expected to take place in anisotropic models irrespective of the type of long-range interaction among particles (see, e.g., \citealt{2011MNRAS.414.3298N,2017MNRAS.468.2222D}), the specific details of its dependence on, for instance, total mass, density slope, anisotropy profile or, in the case of multi-component models, the amount of anisotropy of the individual components (\citealt{1992MNRAS.255..561C}), are far from being fully understood. Numerical studies have argued that cosmological background expansion (\citealt{1995ApJ...440....5C}), compact isotropic inner cores (\citealt{2006ApJ...637..717T}), and instability of the planar orbits of triaxial transient states (\citealt{2007ApJ...670.1027A,2009ApJ...704..372B}) all act as different channels that hamper ROI. Moreover, it is also unclear how the presence of a non-vanishing total angular momentum acts in enhancing or mitigating the instability. While the study of the linear and non-linear stability properties of spherical and spheroidal systems is a classic problem in the fluid dynamical context (see, e.g., \citealt{1961hhs..book.....C} or \citealt{1978trs..book.....T}), much less is known in the context of kinetic theory for equilibria defined as solutions of the collisionless Boltzmann equation (CBE).\\
\indent In this investigation, we extend the numerical analysis of the stability of rotating anisotropic models to a family of cuspy spherical systems with phase space distribution functions of the class introduced by \cite{1979SvAL....5...42O} and \cite{1985AJ.....90.1027M}, modified to impose a net global angular momentum controlled by a single dimensionless parameter $\alpha$, in the same fashion as \cite{1999MNRAS.305..859A} and \cite{2019MNRAS.487..711R}.\\
\indent The article is structured as follows: in Sect.~\ref{models} we outline the construction of spherical, radially anisotropic equilibrium models, and we discuss how to introduce a variable degree of net rotation. In Sect.~\ref{numeric}, we describe the numerical methods, namely the code, the algorithm adopted to generate the initial conditions, and the diagnostics used to characterise the evolution of the models. In Sect.~\ref{sec:results} we present the survey of $N$-body simulations, starting with the scaling tests conducted to assess any dependence on the number of particles, and we analyse the evolution of the morphology, of the density modes, and of the phase space distribution of the models. Finally, in Sect.~\ref{sec:conclusions}, we discuss our results and their astrophysical implications, while Appendix~\ref{sec:cored} compares them with previous stability analyses of cored models.
\section{Models}\label{models}
\subsection{Radially anisotropic systems}\label{mod1}
To investigate the interplay between rotation and velocity anisotropy in spherical systems with a central density cusp, we adopt as a reference the \cite{1990ApJ...356..359H} density profile
\begin{equation}
\label{eq:hernquist}
\rho(r)=\frac{Mr_c}{2\pi r(r+r_c)^3},
\end{equation}
which is characterised by total mass $M$, scale radius $r_c$, and a central density cusp $\rho\propto r^{-1}$. The associated radial mass profile and potential are given by
\begin{equation}\label{eq:mr}
M(r)=M\left(\frac{r}{r+r_{c}}\right)^{2}
\end{equation}
and
\begin{equation}\label{eq:phihernquist}
\Phi(r)=-\frac{GM}{r+r_c}.
\end{equation}
Several equilibria defined by phase space distribution functions with variable degrees of velocity anisotropy are available (see, e.g., \citealt{1987MNRAS.224...13D,1991MNRAS.253..414C,1993MNRAS.261..283L,1995MNRAS.275.1017C,2004AIPC..703..322C,2010MNRAS.401.1091C,2023MNRAS.525.1795B,2024MNRAS.531.5097B}). Here, we focus only on radially anisotropic models defined by an Osipkov-Merritt (hereafter, OM; \citealt{1979SvAL....5...42O,1985AJ.....90.1027M}) distribution function expressed in integral form by using an extension of the \cite{1916MNRAS..76..572E} inversion
\begin{equation}\label{eq:OM}
f(Q)=\frac{1}{\sqrt{8}\pi^2}\int_Q^{0}\frac{{\rm d}^2\rho_a}{{\rm d}\Phi^2}\frac{{\rm d}\Phi}{\sqrt{\Phi-Q}}, 
\end{equation}
\begin{figure}
    \centering
    \includegraphics[width=0.95\columnwidth]{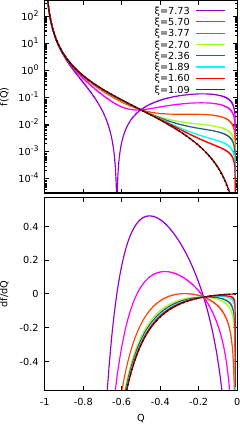}
    \caption{Osipkov-Merritt phase space distribution function for Hernquist models with different values of the anisotropy index $\xi$ (top panel) and its derivative (bottom panel). In both panels, the black dotted line denotes the isotropic case ($\xi=1$ and $Q=\mathcal{E}$).}
    \label{fig:dfOM}
\end{figure}
where
\begin{equation}
Q=\mathcal{E}+{J^2}/{2r_a^2}, 
\end{equation}
 and $\mathcal{E}=v^2/2+\Phi(r)$ and $J$ are the energy and angular momentum of the particle per unit mass; $\rho_a$ is the augmented density, defined by 
 \begin{equation}
 \rho_a(r)\equiv\left(1+{r^2}/{r_a^2}\right)\rho(r),
 \end{equation}
 where the anisotropy radius $r_a$ is a parameter that controls the radial extent of the velocity anisotropy. In practice, at $r\ll r_a$, the model is nearly isotropic, while for $r\gg r_a$ the velocity distribution is dominated by low-angular momentum, nearly radial orbits.\\
\indent For the purely isotropic case (i.e., $r_a\to\infty $ and $Q=\mathcal{E}$), we recover an ergodic distribution function $f(\mathcal{E})$ that, for our choice of $\rho$ and $\Phi$, reads (see also \citealt{2002A&A...393..485B})
\begin{equation}\label{eq:fE}
f(\mathcal{E})=\frac{M\left[3\sin^{-1}q+q(1-q^2)^{1/2}(1-2q^2)(8q^4-8q^2-3)\right]}{8\sqrt{2}\pi^3r_c^3v_c^3(1-q^2)^{5/2}},
\end{equation}
where we set $q=\sqrt{-r_c\mathcal{E}/GM}$ and $v_c=\sqrt{GM/r_c}$. 

\begin{figure}
    \centering
    \includegraphics[width=0.95\columnwidth]{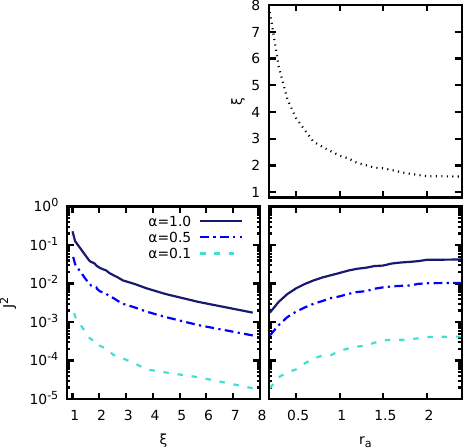}
    \caption{Anisotropy index $\xi$ as a function of $r_a$ (top right panel) and maximal value of the squared total angular momentum $J^2$ as a function of the anisotropy index and anisotropy radius (bottom left and bottom right panels) for Hernquist equilibria with $\alpha=1$ (solid line), 0.5 (dot-dashed line), and 0.1 (dashed line).}
    \label{fig:J2xira}
\end{figure}

For the general anisotropic case with a finite value of $r_a$, the phase space distribution can be written as 
\begin{equation}\label{eq:fqhernquist}
f(Q)=f(\tilde{q})+\frac{M\tilde{q}(1-2\tilde{q}^2)}{\sqrt{2}\pi^3r_cr_a^2v_c^3},
\end{equation}
where $\tilde{q}=\sqrt{-r_cQ/GM}$ and $f(\tilde{q})$ has the same functional form as in Eq.~(\ref{eq:fE}). In practice, Eq.~(\ref{eq:fqhernquist}) is of the form $f(Q)=f(\mathcal{E})+f_a(Q)r_c^2/r_a^2$, where the anisotropic part $f_a(Q)$ is independent of the specific value of the anisotropy radius.\\
\indent We recall that the global amount of anisotropy of a given model can be quantified by the \mbox{Fridman-Polyachenko-Shukhman} anisotropy index (hereafter, anisotropy index for brevity; for details, see, e.g., \citealt{1981SvA....25..533P,1984sv...bookR....F,2008gady.book.....B})
\begin{equation}\label{eq:xi}
 \xi\equiv\frac{2K_r}{K_t},
\end{equation}
where the radial and tangential kinetic energies are given by
\begin{equation}
K_r=2\pi\int\rho(r)\sigma^2_r(r)r^2{\rm d}r,\quad K_t=2\pi\int\rho(r)\sigma^2_t(r)r^2{\rm d}r,
\end{equation}
as a function of $\sigma_r$ and $\sigma_{t}$, i.e., the local values of the radial and tangential components of the velocity dispersion tensor. We adopt the definition of the anisotropy profile
\begin{equation}\label{eq:beta}
\beta(r)=1-\frac{\sigma_t^2(r)}{2\sigma_r^2(r)},
\end{equation}
that, for the specific case of OM models, becomes
\begin{equation}\label{eq:betaOM}
\beta_{\rm OM}(r)=\frac{r^2}{r^2+r_a^2}.
\end{equation}
OM models admit a minimum (critical) value of the anisotropy radius $r_{a,{\rm crit}}$, below which $f(Q)$ is no longer consistent (i.e., it becomes negative somewhere); smaller values of $r_{a,{\rm crit}}$, and hence larger degrees of radial anisotropy, are attainable for steeper density profiles. This critical radius is given by
\begin{equation}
r_{a,{\rm crit}}^2={\rm sup}\left[-\frac{f_a(Q)}{f(\mathcal{E})}r_c^2\right]; \quad Q\in Q^-,
\end{equation}
where $Q^-$ is the range of $Q$ values over which $f_a(Q)\leq 0$.
For the case of anisotropic Hernquist models, the limiting value of the anisotropy radius is $r_{a,{\rm crit}}\approx0.202r_c$ (\citealt{1996ApJ...471...68C,1999ApJ...520..574C}); the most radially anisotropic model simulated in this work, with $\xi=7.73$, lies just inside this limit. This limit should not be confused with the stability threshold $\xi_s$ to which we shall refer in the following. The former is a property of the model family alone, fixed by the requirement that $f(Q)$ be nowhere negative, and carries no implication as to the dynamical stability of the resulting equilibrium. The latter is instead an empirical quantity, established from $N$-body experiments, below which radially anisotropic spherical models are found to preserve their initial symmetry: \cite{1984sv...bookR....F} quote $\xi_s\approx1.70\pm0.25$ (see also \citealt{1996ApJ...471...68C}), \cite{2002MNRAS.332..901N} find $1.6\lesssim\xi_s\lesssim1.8$ from $N$-body simulations of OM models, including the Hernquist case, while \cite{2020MNRAS.494.1027D} obtain $\xi_s\simeq1.50\pm0.20$ for a family of OM models with different inner density slopes; the value $\xi_s\approx1.60$ recovered from our own survey (see Sect.~\ref{sec:scaling}) is consistent with all of them. We note that, for models whose distribution function diverges for $J\to0$, the analyses of \cite{1987MNRAS.224.1043P}, \cite{palmer1994stability} and \cite{2019MNRAS.487..711R} indicate that a spectrum of unstable modes is present at any degree of anisotropy, so that no sharp threshold exists. Such a divergence occurs in the cored models discussed in Appendix \ref{sec:cored}, but not in the OM models considered here, whose $f(Q)$ remains finite for $J\to0$. In the following, $\xi_s$ is nonetheless to be understood as a practical, resolution-dependent indicator rather than as a sharp boundary, since it is determined empirically from the simulations. The distribution functions of selected OM anisotropic Hernquist models are represented, together with their derivatives, in Fig.~\ref{fig:dfOM}. Note that the distribution functions of equilibria characterised by large values of $\xi$ are markedly non-monotonic, as illustrated also by the sign changes in ${\rm d}f/{\rm d}Q$. For this reason, their stability cannot be ensured by the \cite{1961SvA.....4..859A} criterion, according to which a spherical system with an ergodic distribution function $f(\mathcal{E})$ such that ${\rm d}f/{\rm d}\mathcal{E}<0$ is stable (see, e.g., \citealt{2008gady.book.....B}). We recall that the monotonicity of $f$ is a sufficient, but not necessary, condition for stability: its violation does not imply that a model is unstable. Moreover, when $f$ depends on $J$ as well as on $\mathcal{E}$ (in our case through the auxiliary variable $Q$), the condition $\partial f/\partial\mathcal{E}<0$ at fixed $J$ guarantees stability against radial perturbations only (\citealt{1976A&A....50..467G,2011CMaPh.302..161L}). Since $\partial f/\partial\mathcal{E}|_J={\rm d}f/{\rm d}Q$ for OM models, the non-monotonicity of $f(Q)$ at large $\xi$ implies that not even this weaker guarantee is available for the models considered here. The relations between the total angular momentum and the key parameters $\xi$ and $r_a$ are depicted in Fig.~\ref{fig:J2xira}.
\subsection{Introducing rotation}\label{rotscheme}

\indent As in recent investigations by \cite{2019MNRAS.487..711R}, \cite{2021MNRAS.502.4762B}, and an earlier study by \cite{1996MNRAS.280..689P}, we define a family of rotating equilibria obtained by deploying the so-called ``\cite{1960MNRAS.120..204L} daemon''. Starting from a parent, non-rotating OM distribution function $f_0(Q)$, we introduce the explicit dependence on $J_z$ by augmenting it as
\begin{equation}\label{eq:falpha}
f(Q,\alpha)=f_0(Q)+\alpha f_0(Q){\rm Sign}(J_z),
\end{equation}
where $|\alpha|\leq 1$ is the parameter that controls the amount of global angular momentum and ${\rm Sign}(x)$ denotes the sign function. By construction, one has
\begin{equation}
\int_{-J}^{+J}{\rm Sign}(J_z){\rm d}J_z=0,
\end{equation}
therefore, such an augmentation\footnote{We note that the distribution function defined by Eq.~(\ref{eq:falpha}) features by construction a singular gradient, as it involves a discontinuous ${\rm Sign}(x)$ function. For this reason, it is not likely to fulfil \citeauthor{1961SvA.....4..859A}'s \citeyearpar{1961SvA.....4..859A} Fourth Law for stability. In principle, using a smooth function of $J$ in lieu of the sign function could lead to more tractable models.} of the phase space distribution does not alter the degree of anisotropy of the parent OM model $f_0(Q)$. Instead, for an orbit corresponding to a given value of $Q$, the probability of having $J_z>0$ ($<0$) is multiplied by $1+\alpha$ ($1-\alpha$) with respect to the non-rotating case. For simplicity, we consider only models with counterclockwise net rotation (i.e., total $J_z>0$), hence $\alpha\in[0,1]$; the case $\alpha=1$ corresponds to a maximally rotating model, while $\alpha=0$ to the parent, non-rotating equilibrium defined by $f(Q,\alpha)=f_0(Q)$. We also note that the anisotropy index $\xi$ does not have any explicit dependence on the direction of the components of the tangential velocity $v_t=\sqrt{v_\vartheta^2+v_\varphi^2}$. Moreover, the anisotropy profile $\beta$ defined in Eqs.~(\ref{eq:beta}) and (\ref{eq:betaOM}) remains unaffected too. Interestingly, this is not the case if the \cite{1960MNRAS.120..204L} daemon is applied to other classes of non-rotating anisotropic equilibria, such as the ones proposed by \cite{1987MNRAS.224...13D}. The degree of anisotropy of models whose azimuthal and polar velocity dispersions $\sigma_\varphi$ and $\sigma_\vartheta$ are characterised by different functional forms is indeed modified by this augmentation, because the radial profile of the tangential component of the velocity dispersion is defined by $\sigma_t^2=\sigma_\varphi^2+\sigma_\vartheta^2$ (see, e.g., \citealt{2019MNRAS.487..711R}). 

For general interest, we note that the explicit construction (i.e., by means of a further generalisation of the Eddington inversion protocol, see, e.g., \citealt{1986PhR...133..217D,Lacroix_2018}) of analytical, self-consistent models with phase space distributions $f(Q,J_z)$ simultaneously characterised by velocity anisotropy and non-vanishing total angular momentum is a rather cumbersome task. We recall that, for any spherically symmetric or axisymmetric potential, orbits with frequency $\varpi$ around the $z$-axis conserve the Jacobi constant in an inertial frame
\begin{equation}\label{eq:jacobi}
C_{\rm J}=2\varpi (xv_y-yv_x)-2\Phi(x,y,z)-(v_x^2+v_y^2+v_z^2)=2\varpi J_z-2\mathcal{E}.
\end{equation}
Formally, it is possible to extend the OM parametrisation by taking advantage of the Jacobi constant $\tilde{C}_{\rm J}=2\varpi J_z-2Q$, which, expressed in spherical coordinates, becomes
\begin{equation}\label{eq:jacobisph}
\tilde{C}_{\rm J}=2\varpi r\sin\vartheta v_\varphi-[v_r^2+(1+r^2/r_a^2)(v_\vartheta^2+v_\varphi^2)]-2\Phi(r),
\end{equation}
where $r\sin\vartheta=R=\sqrt{x^2+y^2}$ denotes the cylindrical radius and $v_r$ and $v_t$ are the radial and tangential components of the velocity vector. Following \citet[Eqs.~7--10]{1985AJ.....90.1027M}, the formal integration of $f(\tilde{C}_{\rm J})$ should be performed in ${\rm d}v_r{\rm d}v_\vartheta{\rm d}v_\varphi$, rather than ${\rm d}v_r{\rm d}v_t$ as in Eq.~(\ref{eq:jacobisph}), because the two components of the tangential velocity are decoupled. By doing so, one is left with an axisymmetric augmented density which is associated with a double integral inversion formula. The problem is now formally equivalent to the construction of a two-integral distribution function for an effective model with cylindrical symmetry. The latter can be evaluated via the \cite{1993MNRAS.262..401H} technique, which requires performing contour integrals in the complex plane; unfortunately, such a task is computationally demanding even for simple parent equilibria.\\

\section{Numerical methods}\label{numeric}
\subsection{The code}\label{code}
The $N$-body simulations performed in this investigation have been carried out using the scalar version of the publicly available {\sc fvfps} code introduced in \citet[see also \citealt{2002MNRAS.332..901N}]{2003MSAIS...1...18L}. {\sc fvfps} uses a {\sc fortran} implementation of the \cite{1986Natur.324..446B} tree scheme for the force evaluation combined with the \cite{2002JCoPh.179...27D} fast multipole method (see also \citealt{2011EPJP..126...55D}). To conduct appropriate scaling tests, the simulations have a total number of particles $N$ ranging between $3\times10^3$ and $10^6$; all particles have the same mass $m_i=M/N$. \\
\indent In the evaluation of the gravitational forces, we set the opening angle of the tree scheme to $\theta_{\rm min}=0.5$. The forces between neighbouring particles below a cut-off distance are smoothed via cubic splines (\citealt{2001MNRAS.324..273D}). In this work, such a softening length is in the range $0.002\leq\epsilon_{\rm soft}\leq 0.05$, with the extremes corresponding to $N=10^6$ and $N=3\times 10^3$.\\
\indent We adopt a coordinate system such that positions and velocities are expressed in units of $r_c$ and $v_c$ (introduced in Sect.~\ref{mod1}), normalised such that $G=M=r_c=1$. The relevant equations of motion are solved with a second-order Verlet scheme. The timestep $\Delta t$, which is given in units of the dynamical time
\begin{equation}\label{eq:tdyn}
T_{\rm dyn}=\sqrt{\frac{r_c^3}{GM}},
\end{equation}
is allowed to vary according to the criterion (see, e.g., \citealt{2011EPJP..126...55D})
\begin{equation}\label{eq:dt}
\Delta t \equiv \frac{\eta}{\sqrt{{\rm max}\nabla^2\Phi}},
\end{equation}
where $\eta$ is a dimensionless control parameter of order $10^{-2}$. For example, with $\eta=0.01$ and $\epsilon_{\rm soft}=0.03$, we obtain that $0.006 \lesssim\Delta t\lesssim 0.013$ for $N=10^5$. We verified that the error on energy conservation over $t=300T_{\rm dyn}$ is around one part in $10^6$, while the centre of mass of the models initially set at the origin of the frame has a drift of about $10^{-4}$ in units of $r_c$ for $N\approx10^5$.
\subsection{Initial conditions}\label{incond}
The initial conditions for the $N$-body simulations are implemented as follows. First, the radial coordinates $r_i$ of the $N$ equal-mass particles are sampled with a one-dimensional Monte Carlo scheme from the cumulative mass distribution obtained by inverting the radial mass profile expressed in Eq.~(\ref{eq:mr}). The angular coordinates $\vartheta$ and $\varphi$ are then randomly assigned by sampling $\cos\vartheta$ from a uniform distribution in $(-1,1)$, and $\varphi$ from a uniform distribution in $(0,2\pi)$, while the potential $\Phi(r_i)$ is evaluated by computing Eq.~(\ref{eq:phihernquist}) at the position of each particle.\\
\indent The OM phase space distribution function $f(Q)$ is computed according to Eq.~(\ref{eq:fqhernquist}) and saved on a staggered mesh with finer grain around the steeper slope regions, where $Q\to -1$ and $Q\to 0$. Here, we typically use $5\times 10^3$ mesh points. Following \cite{2002MNRAS.332..901N} and \cite{2015JPlPh..81e4904D}, we construct the dimensionless vector ${\mathbf u}=(u_1,u_2,u_3)$, with components sampled from a uniform distribution in $(-1,1)$, rejecting the triplets with $||\mathbf{u}||>1$. The putative physical velocity components are then defined as
\begin{equation}\label{eq:vputative}
v_r=\sqrt{2|\Phi|}u_1,\quad v_\vartheta=\frac{\sqrt{2|\Phi|}u_2}{\sqrt{1+r^2/r_a^2}},\quad v_\varphi=\frac{\sqrt{2|\Phi|}u_3}{\sqrt{1+r^2/r_a^2}},
\end{equation}
where, for simplicity, we dropped the suffix $i$. At this point, Eq.~(\ref{eq:fqhernquist}) is evaluated for $Q=(1-u^2)\Phi(r_i)$ and $Q=\Phi(r_i)$ using a linear interpolation. According to the von Neumann rejection method, a random number $\zeta$ is extracted from a uniform distribution in $(0,1)$ and, if $f(Q)\geq\zeta f(\Phi)$, then the velocity vector $(v_r,v_\vartheta,v_\varphi)$ is accepted and position and velocity are converted to Cartesian coordinates; otherwise the velocity is discarded, and the procedure is repeated.\\
\indent Up to now, this sampling protocol does not have any control on the direction and norm of the total angular momentum $\mathbf{J}=(J_x,J_y,J_z)$ of the system. While for small values of $N$ two independent samples of a given $f(Q)$ might have considerably different total angular momenta, for large system sizes (say $N\gtrsim 8\times 10^4$), the total angular momentum of each realisation is nearly vanishing. To introduce a net rotation to the OM models, following \cite{1999MNRAS.305..859A} and \cite{2019MNRAS.487..711R}, we take $v_\varphi=|v_\varphi|$ either in the velocity component given in Eq.~(\ref{eq:vputative}) or at the end of the velocity sampling. We then extract a random number $p_i$ from a uniform distribution in $[0,1]$ and, if $p_i\leq(1-\alpha)/2$, we set the new velocity component to $v_\varphi=-|v_\varphi|$. In this way a fraction $(1+\alpha)/2$ of the particles retains a prograde azimuthal velocity and a fraction $(1-\alpha)/2$ is made retrograde, in accordance with the probabilities implied by Eq.~(\ref{eq:falpha}): for $\alpha=0$ the two fractions are equal and the realisation carries no net angular momentum, while for $\alpha=1$ all particles are prograde and the rotation is maximal.\\
\begin{figure*}
    \centering
    \includegraphics[width=0.95\textwidth]{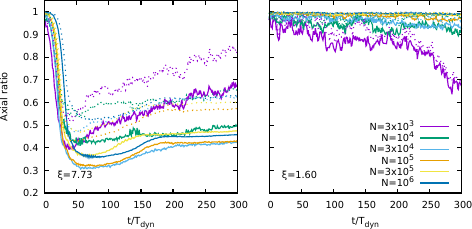}
    \caption{Time evolution of the axial ratios $c/a$ (solid lines) and $b/a$ (dotted lines) for the maximally anisotropic $N$-body model ($\xi\approx 7.73$, left panel) and a nearly isotropic $N$-body model ($\xi\approx 1.60$, right panel) for different total numbers of particles $3\times 10^3\leq N\leq 10^6$. In both cases, models do not have net rotation (i.e., $\alpha=0$).}
    \label{fig:testscaling}
\end{figure*}
\indent To have the total angular momentum aligned along one of the axes (say the $z$-axis), we rotate the coordinate system by an angle $\omega=\cos^{-1}\mathbf{J}\cdot\hat{z}/J_{\rm tot}$, where $J_{\rm tot}\equiv|\mathbf{J}|$, around the axis $\mathbf{n}$$=(J_y/\sqrt{J_x^2+J_y^2},-J_x/\sqrt{J_x^2+J_y^2},0)$. The associated rotation matrix $\mathcal{R}$ is given by
\begin{equation}
\mathcal{R}=\mathbb{I}+\sin\omega\mathcal{K}+(1-\cos\omega)\mathcal{K}\mathcal{K},
\end{equation}
where $\mathbb{I}$ is the $3\times 3$ identity matrix and $\mathcal{K}$ is defined as a function of the three components of the rotation axis $n_1$, $n_2$, and $n_3$ as
\begin{equation}
\mathcal{K}=\begin{pmatrix}
0 & -n_3 & n_2\\
n_3 & 0 & -n_1\\
-n_2 & n_1 & 0
\end{pmatrix}	
.
\end{equation}
In Fig.~\ref{fig:J2xira}, we show the maximal value of $J_{\rm tot}^2$ (labelled $J^2$ in the figure) that can be attained in an $N$-body realisation as a function of the anisotropy index (bottom left panel) and anisotropy radius (bottom right panel), for cases with $\alpha=1$, 0.5 and 0.1. As expected, for increasingly radially anisotropic parent OM models, the rotating $N$-body systems can support smaller amounts of total angular momentum, regardless of the specific value of $\alpha$. Another indicator of global rotation is the so-called \cite{1971A&A....11..377P} rotation parameter, defined for a system of mass $M$, total energy $\mathcal{E}_{\rm tot}$, and angular momentum $J_{\rm tot}$ as 
\begin{equation}\label{eq:peebles}
\lambda=\frac{J_{\rm tot}\sqrt{-\mathcal{E}_{\rm tot}}}{GM^{5/2}}.
\end{equation}
For our case of a Hernquist sphere, in units such that $G=M=r_c=1$, we have $\mathcal{E}_{\rm tot}=-1/12$ (\citealt{2021A&A...653A.140B}) and thus $\lambda=J_{\rm tot}\sqrt{1/12}\approx 0.29J_{\rm tot}$. In principle, different combinations of $\alpha$ and $\xi$ can yield the same $J_{\rm tot}$ (see bottom left panel of Fig.~\ref{fig:J2xira}), thus implying the same value of the rotation parameter $\lambda$.
\subsection{Diagnostics}\label{diagn}
In line with previous ROI studies, we analyse the departure of the system from the initial spherical symmetry, as quantified by its axial ratios $c/a$ and $b/a$, where $c\leq b\leq a$ are the three semi-axes of the system of particles. At every dynamical time (roughly corresponding to an interval of 100 $\Delta t$), we evaluate the second-order tensor
\begin{equation}\label{eq:tensor}
\mathcal{I}_{ij}\equiv\frac{M}{N}\sum_{k=1}^N r_i^{(k)}r_j^{(k)} 
\end{equation}
related to the inertia tensor of the system by the relation $I_{ij}\equiv{\rm Tr}(\mathcal{I})\delta_{ij}-\mathcal{I}_{ij}$. Equation (\ref{eq:tensor}) is computed for the particles within the Lagrangian radii $r_{10\%}$, $r_{50\%}$, $r_{70\%}$ and $r_{90\%}$, which enclose 10\%, 50\%, 70\% and 90\% of the total mass of the system.\\ 
\indent The matrix $\mathcal{I}_{ij}$ is diagonalised iteratively (following the scheme of \citealt{2002MNRAS.332..901N}, see their Appendix A), with the requirement that the percentage difference of the largest eigenvalue between two iterations does not exceed $10^{-3}$. On average, for $N\approx 10^5$ the process requires about 10 iterations. Once the three eigenvalues $\mathcal{I}_1\geq \mathcal{I}_2\geq \mathcal{I}_3$ of $\mathcal{I}_{ij}$ are recovered, for a heterogeneous ellipsoid of semi-axes $a,b$ and $c$, we have $\mathcal{I}_1=Aa^2$, $\mathcal{I}_2=Ab^2$ and $\mathcal{I}_3=Ac^2$, where $A$ is a constant depending on the density profile. The axial ratios 
are then defined by $b/a=\sqrt{\mathcal{I}_2/\mathcal{I}_1}$ and $c/a=\sqrt{\mathcal{I}_3/\mathcal{I}_1}$, so that the ellipticities in the principal planes are $\epsilon_1=1-\sqrt{\mathcal{I}_2/\mathcal{I}_1}$ and $\epsilon_2=1-\sqrt{\mathcal{I}_3/\mathcal{I}_1}$. We define systems with $c/a\sim b/a\lesssim 0.5$ as prolate, with $c/a\sim b/a\gtrsim 0.5$ as oblate, while systems having $b/a>0.5$ and $c/a<0.5$ are triaxial.\\
\indent A finer analysis of the evolution of systems prone to instabilities can be performed by assessing the growth of the modes of their density profile. It is convenient to define a spectral decomposition of order $n$ by expressing the density in cylindrical coordinates $(R,z,\varphi)$ as follows
\begin{equation}\label{eq:Cn}
C_n(t)=\frac{1}{2\pi}\int_0^{+\infty}\int_{-\infty}^{+\infty}\int_0^{2\pi}\rho(R,z,\varphi,t)e^{-\mathrm{i} n\varphi}R{\rm d}\varphi{\rm d}z{\rm d}R.
\end{equation}
\begin{figure*}
    \centering
    \includegraphics[width=0.95\textwidth]{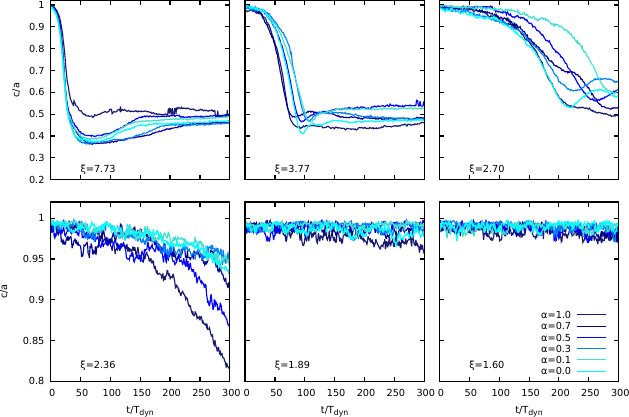}
    \caption{Time evolution of the axial ratio $c/a$ of $N$-body models with initial anisotropy index $\xi\approx7.73$, 3.77, 2.70, 2.36, 1.89, and 1.60 (from left to right and top to bottom). Darker shades of blue indicate higher values of the rotation parameter $\alpha$, corresponding to higher values of the total angular momentum.}
    \label{fig:ca}
\end{figure*}
We then define the normalised quantities $A_n(t)=C_n(t)/C_0(t)$, whose norms $|A_n(t)|$ are proportional to the amplitude of the growing modes of $\rho$. Assuming exponential growth (at least in the first phase after the onset of the instability), the growth rate is given by the logarithmic derivative
\begin{equation}\label{eq:gamma}
\gamma_n=\frac{{\rm d}\log|A_n(t)|}{{\rm d}t},
\end{equation}
with associated pattern frequency
\begin{equation}\label{eq:omega}
\Omega_n=\frac{{\rm d}{\rm Arg}[A_n(t)]}{{\rm d}t},
\end{equation}
where ${\rm Arg}(\mathbf{z})=\arctan[{\rm Im}(\mathbf{z})/{\rm Re}(\mathbf{z})]$ for ${\rm Re}(\mathbf{z})>0$ and $\pi+\arctan[{\rm Im}(\mathbf{z})/{\rm Re}(\mathbf{z})]$ for ${\rm Re}(\mathbf{z})<0$.
\begin{figure}
    \centering
    \includegraphics[width=\columnwidth]{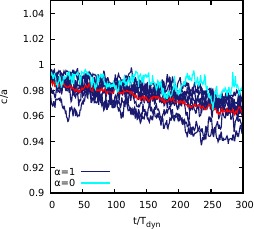}
    \caption{Time evolution of the axial ratio $c/a$ of $N$-body models with initial anisotropy index $\xi\approx1.89$ for the stable $\alpha=0$ (cyan line) case and different realisations with $\alpha=1$ (dark blue lines) for the rotating unstable case. The red line marks the average.}
    \label{fig:realization}
\end{figure}
\begin{figure}
    \centering
    \includegraphics[width=\columnwidth]{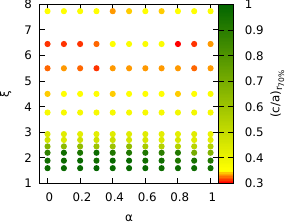}
    \caption{Final axial ratio of all $N$-body models, illustrated as a function of the initial anisotropy index $\xi$ and rotation parameter $\alpha$.}
    \label{fig:caalphaxi}
\end{figure}

\begin{figure}
    \centering
    \includegraphics[width=\columnwidth]{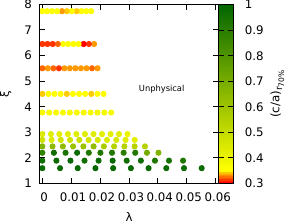}
    \caption{Final axial ratio of all $N$-body models, illustrated as a function of the initial anisotropy index $\xi$ and Peebles parameter $\lambda$.}
    \label{fig:lambda}
\end{figure}

We numerically compute the integrals appearing in Eq.~(\ref{eq:Cn}) after having evaluated the integrand at a given simulation snapshot on a cylindrical grid of $N_g=64\times32\times32$ nodes in $(R,z,\varphi)$. Alternatively, $C_n(t)$ can also be obtained from the particles' positions by substituting the integrals over $\rho(R,z,\varphi)e^{-\mathrm{i} n\varphi}R$ with a sum of delta functions as
\begin{equation}
C_n(t)=\frac{1}{2\pi N}\sum_{i=1}^NR_i(t)e^{-\mathrm{i}n\varphi_i},
\end{equation}
in a similar fashion to what was done in \cite{2022MNRAS.512.3015B} to compute the evolution of the Fourier modes of a non-equilibrium phase space distribution defined for the $N$-body systems as
\begin{equation}\label{eq:dfdiscrete}
f(\mathbf{r},\mathbf{v})\equiv \frac{1}{N} \sum_{i=1}^N \delta^3(\mathbf{r}-\mathbf{r}_i)\delta^3(\mathbf{v}-\mathbf{v}_i).
\end{equation}
With such a definition, the Fourier mode associated with a given $\mathbf{k}=(\mathbf{k}_\mathbf{r},\mathbf{k}_\mathbf{v})$ becomes
\begin{equation}\label{eq:fourierdiscr}
    f^*(\mathbf{k},t)=\frac{1}{N}\sum_{i=1}^N e^{\mathrm{i}(\mathbf{k}_\mathbf{r} \cdot \mathbf{r}_i(t) +\mathbf{k}_\mathbf{v} \cdot \mathbf{v}_i(t))},
\end{equation}
where $\mathbf{r}_i(t)$ and $\mathbf{v}_i(t)$ are the time-dependent positions and velocities of a particle of index $i$, and, following \cite{2022MNRAS.512.3015B}, a Fourier mode is labelled by the modulus $k\equiv\|\mathbf{k}\|=(\|\mathbf{k}_\mathbf{r}\|^2+\|\mathbf{k}_\mathbf{v}\|^2)^{1/2}$ of its wave vector.
\section{Simulations and results}\label{sec:results}
\subsection{Scaling with the number of particles}\label{sec:scaling}
\begin{figure*}
    \centering
    \includegraphics[width=0.97\textwidth]{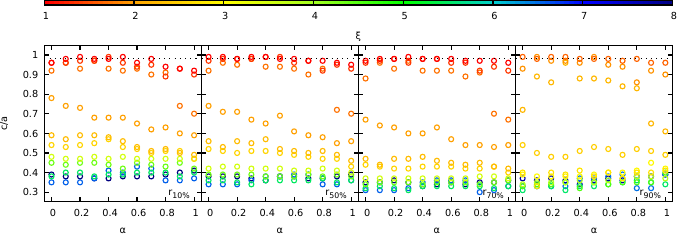}
    \includegraphics[width=0.97\textwidth]{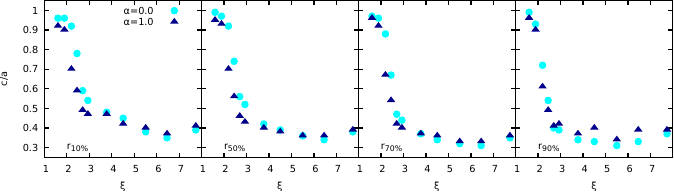}
 \includegraphics[width=0.97\textwidth]{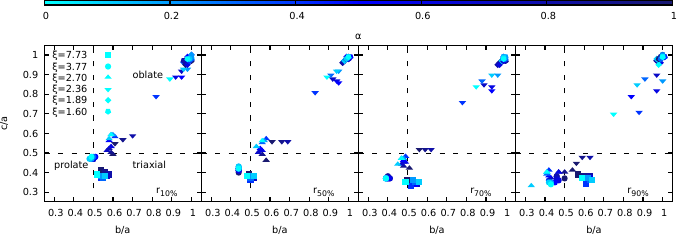}
   \caption{Top panels: Final values of the axial ratios $c/a$ of selected $N$-body models, depicted as a function of the rotation parameter $\alpha$ evaluated within (from left to right) $r_{10\%}$, $r_{50\%}$, $r_{70\%}$ and $r_{90\%}$. The initial values of the anisotropy index of the $N$-body models are illustrated by the colour map. Middle panels: Final values of the axial ratio $c/a$ as a function of the anisotropy index $\xi$ for non-rotating (cyan circles) and maximally rotating (blue triangles) $N$-body models. Bottom panels: Final values of the axial ratios $c/a$ and $b/a$ evaluated within $r_{10\%}$, $r_{50\%}$, $r_{70\%}$ and $r_{90\%}$ for $N$-body models with initial values of the anisotropy index ranging from maximally anisotropic ($\xi = 7.73$) to nearly isotropic conditions ($\xi = 1.60$). The value of the rotation parameter $\alpha$ is colour-coded in shades of blue.}
    \label{fig:calagrangian}
\end{figure*}
In line with \citet[see also \citealt{1997ApJ...490..136M}]{2020MNRAS.494.1027D}, we first performed a suite of scaling tests by evolving models (with selected choices of $\xi$ and $\alpha$) sampled with values of $N$ ranging from $3 \times 10^3$ to $10^6$. As a reference observable, we measured the evolution of the triaxiality of the models, as it gives an intuitive insight into the onset of any instabilities. In agreement with previous results, we find that extremely radially anisotropic systems (say $\xi\gtrsim 4$) evolve to progressively more non-spherical end states as $N$ increases. In the low-$N$ cases, ROI, albeit faster, is stopped by discreteness effects once the minimal axial ratio $c/a$ reaches $\approx 0.4$ and the departure from spherical symmetry diminishes. For initial values of $\xi$ close to $\xi_s$, the behaviour is the opposite: models with smaller $N$ become flatter than their higher-$N$ counterparts with the same value of $\xi$. Such a behaviour manifests itself in both rotating and non-rotating models; therefore, for brevity, in Fig.~\ref{fig:testscaling} we illustrate the outcome of the scaling tests for two representative non-rotating cases: a maximally anisotropic Hernquist model with $\xi\approx 7.73$, close to the consistency limit introduced in Sect.~\ref{mod1}, and a marginally stable model with $\xi\approx 1.60\simeq\xi_s$.
   Consistent with previous numerical studies (see, e.g., \citealt{2017MNRAS.468.2222D}), highly ROI-prone systems are more markedly triaxial, while marginally stable models are approximately prolate, with considerable departure from spherical symmetry already emerging at $t\approx 50T_{\rm dyn}$ for $N\lesssim 10^4$.\\
\indent No significant $N$-dependent variations in either the evolution or the final structure of the unstable models have been noticed for $N\gtrsim 10^5$; we therefore adopt $N=10^5$ as the reference particle number for the main survey of simulations.
\begin{figure}
    \centering
    \includegraphics[width=0.9\columnwidth]{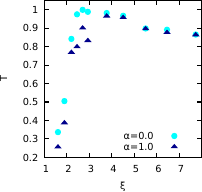}
    \caption{Values of the triaxiality index $T$ as a function of $\xi$ for the final configurations of the models depicted in Fig.~\ref{fig:calagrangian} (middle panels); for all models, the triaxiality index is evaluated within the final half-mass radius.}
    \label{fig:triaxial}
\end{figure}
\subsection{Evolution of the intrinsic morphology}

As a general observation, we note that the growth of a non-spherical perturbation sets in systematically earlier in models characterised by increasing values of the initial anisotropy index $\xi$, irrespective of the specific value of the rotation parameter $\alpha$. Such a behaviour is consistent with the one already observed in non-rotating OM models (see, e.g., Fig.~7 in \citealt{2020MNRAS.494.1027D}). As an example, in Fig.~\ref{fig:ca} we illustrate the evolution up to $t=100T_{\rm dyn}$ of the axial ratio $c/a$ computed within the Lagrangian radius $r_{70\%}$ of models with $\xi=7.73$, 3.77, 2.70, 2.36, 1.89 and 1.60 and $\alpha=0.0,0.1,0.3,0.5,0.7,1.0$, indicated by increasingly darker shades of blue. For a given initial degree of radial anisotropy close to the stability threshold (i.e., $\xi\gtrsim 1.60$), the presence of non-vanishing total angular momentum accelerates the onset of non-spherical perturbations. This effect is not limited to single realisations or the specific value of $N$ used in the simulations. For each combination of $\alpha$ and $\xi$, we ran an additional set of 20 independent realisations. We find that the scatter of $c/a$ with respect to the ensemble-average ranges from $\approx2\%$ to about $10\%$. As an example, in Fig.~\ref{fig:realization} we present, for $\xi=1.89$, the evolution of $c/a$ for the stable non-rotating model of Fig.~\ref{fig:ca} together with six randomly selected realisations of the maximally rotating ($\alpha=1$) model. The rotation induces in all cases a departure from spherical symmetry occurring at different times between $5$ and $300T_{\rm dyn}$. For comparison, we indicate the realisation average with a red line. In addition, higher values of $J_z$ correspond to significantly flatter transient states and, possibly, final configurations.\\
\indent Interestingly, for initial values of $\xi$ close to the consistency limit ($\xi=7.73$), the evolution of $c/a$ appears to be non-monotonic in $\alpha$. For values of $\alpha$ around 0.5, the departure from spherical symmetry leads to lower values of $c/a$ at around $t\approx 50T_{\rm dyn}$, with respect to the non-rotating ($\alpha=0$) and maximally rotating cases ($\alpha=1$). Overall, the presence of rotation has a slight mitigating effect on the ROI for $\alpha\gtrsim0.5$ for parent OM models characterised by a large degree of radial anisotropy. The final values of $c/a$ are displayed for the whole survey of simulations in Fig.~\ref{fig:caalphaxi}. With special reference to high-rotation models (i.e., $\alpha\gtrsim 0.6$), we note that almost isotropic final equilibria are more spherical, while anisotropic ones are more flattened. The final flattening of the simulated systems can also be appreciated as a function of the Peebles rotation parameter $\lambda$ defined in Eq.~(\ref{eq:peebles}) in Fig.~\ref{fig:lambda}. The final flattening is primarily set by $\xi$: at fixed anisotropy, increasing $\lambda$ produces a mild reduction of $c/a$ for $2\lesssim\xi\lesssim3$ and no systematic dependence at larger $\xi$, so that the behaviour discussed above is preserved when the degree of rotation is measured by $\lambda$ in place of $\alpha$. The representation in terms of $\lambda$ makes apparent, however, that the range of rotation accessible to the models narrows markedly as the anisotropy grows: the nearly isotropic configurations reach $\lambda\approx0.06$, whereas the most radially anisotropic ones are confined to $\lambda\lesssim0.02$.\\
\begin{figure}
    \centering
    \includegraphics[width=\columnwidth]{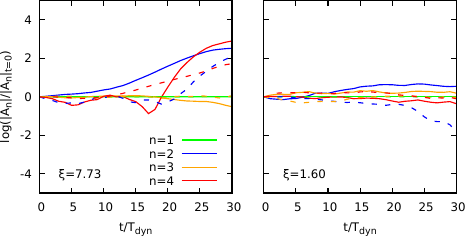}
    \caption{Time evolution of $|A_n|$ (for $n=1$, 2, 3, and 4), normalised to its initial value, for the maximally anisotropic $N$-body models with $\xi \approx 7.73$ (left panel) and the nearly isotropic $N$-body models with $\xi\approx 1.60$ (right panel). The solid lines refer to $N$-body models with no net angular momentum ($\alpha=0$), while the dashed lines refer to those with the maximum rotation ($\alpha=1$); in all cases $N=10^5$.}
    \label{fig:modesn}
\end{figure}
\begin{figure}
    \centering
    \includegraphics[width=\columnwidth]{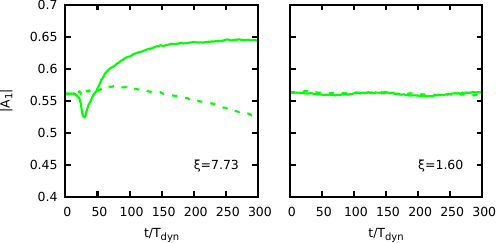}
    \caption{Time evolution of $|A_1|$ in the maximally anisotropic models with $\xi\approx7.73$ (left panel), and the nearly isotropic models with $\xi\approx 1.60$ (right panel). The solid lines refer to $N$-body models with no net angular momentum ($\alpha=0$), while the dashed lines refer to those with the maximum rotation ($\alpha=1$); in all cases $N=10^5$.}
    \label{fig:mode1}
\end{figure}
\begin{figure*}
    \centering
    \includegraphics[width=0.95\textwidth]{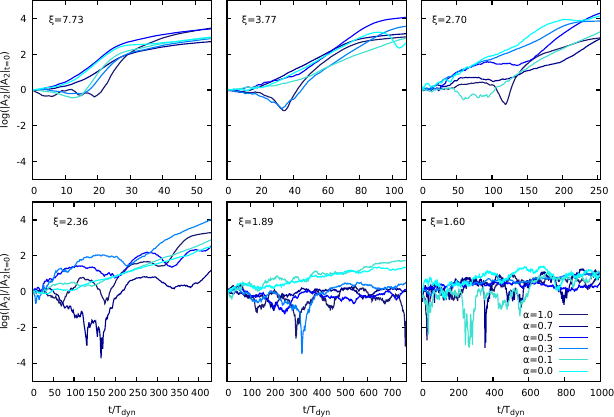}
    \caption{Time evolution of $|A_2|$, normalised to its initial value, for the same models displayed in Fig.~\ref{fig:ca}.}
    \label{fig:mode2}
\end{figure*}
\begin{figure*}
    \centering
    \includegraphics[width=0.29\textwidth]{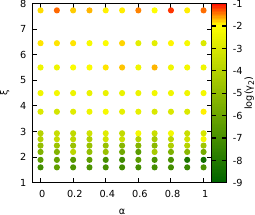}
    \includegraphics[width=0.635\textwidth]{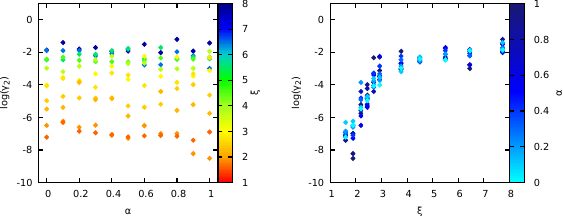}
    \caption{Logarithm of the growth rate $\gamma_2$ of the $n=2$ mode, colour-coded as a function of the initial anisotropy index $\xi$ and rotation parameter $\alpha$ (left panel) and, separately, as a function of the rotation (middle panel) and anisotropy (right panel) parameters.}
    \label{fig:gammaalphaxi}
\end{figure*}
\indent In the top panels of Fig.~\ref{fig:calagrangian} we show the value of $c/a$ at $t=10^3T_{\rm dyn}$ as a function of the rotation parameter $\alpha$ (the time has been chosen to represent a stage at which all numerical models undergoing instabilities are sufficiently relaxed). The various panels refer to the values of the Lagrangian radii within which the three semi-axes of the system have been computed. The general qualitative behaviour is to have increasingly flatter systems at fixed $\alpha$ for increasing anisotropy index $\xi$. Models with $\xi$ near the stability threshold are slightly flatter, with $c/a\approx 0.98$ for larger $\alpha$. The behaviour becomes more evident, albeit with some (numerical) fluctuations, for values of the anisotropy index in the interval $2.3\lesssim\xi\lesssim3.5$. For $\xi\approx3.8$ no appreciable difference in the final values of $c/a$ can be observed in models with different values of $\alpha$. For all values of $\xi$ above $\sim2.4$ explored here, the flattening as a function of $\alpha$ increases when measured within increasing values of the Lagrangian radius, up to $r_{70\%}$; no particular behaviour is observed in $N$-body models starting from less anisotropic initial conditions. In the middle panels of Fig.~\ref{fig:calagrangian} we assess, within the same Lagrangian radii as above, the dependence of the final axial ratio $c/a$ on the initial degree of anisotropy, in the case of the non-rotating ($\alpha=0$) and maximally rotating ($\alpha=1$) $N$-body models. We observe that, irrespective of the Lagrangian radius, the final values of $c/a$ are systematically lower in maximally rotating systems with $2\lesssim\xi\lesssim 3$, comparable to those of the non-rotating ones around $\xi\approx4$, while for the most anisotropic systems ($\xi\gtrsim5$) the behaviour reverses. 

The flavour of the departure from spherical symmetry (i.e., the locus occupied by an $N$-body model in the $c/a$ versus $b/a$ morphological plane) varies strongly with $\xi$ and $\alpha$, as illustrated in the bottom panels of Fig.~\ref{fig:calagrangian}. The final configurations of the maximally anisotropic $N$-body models are the most triaxial across all Lagrangian radial ranges. Below $r_{70\%}$, only the $N$-body models with initial $\xi=3.77$ have a prolate structure, while lower initial degrees of anisotropy are associated with oblate final states. At $r_{90\%}$, due to the less populated outer regions, the final axial ratios have a larger scatter both along and across the diagonal of the morphological plane. In agreement with previous numerical studies, no system is found to be flatter than an E7 model (i.e., $c/a$ is always greater than 0.3). 

Further insight into the morphology of the final configurations of the $N$-body models can be obtained by evaluating the triaxiality index
\begin{equation}\label{eq:triaxial}
T=\frac{a^2-b^2}{a^2-c^2},
\end{equation}
such that $T=0$ ($T=1$) corresponds to an oblate (prolate) axisymmetric spheroid. Intermediate values of $T$ correspond to triaxial systems. We find that, for all values of $\alpha$ explored in this work, the final value of $T$ is a non-monotonic function of the anisotropy index with a maximum at around $\xi\approx3$. In Fig.~\ref{fig:triaxial}, we illustrate $T$ as a function of $\xi$ evaluated within the half-mass radius of the non-rotating ($\alpha=0$) and maximally rotating ($\alpha=1$) $N$-body models. In both cases, while nearly isotropic systems tend to be almost oblate, the triaxiality increases up to $\approx 1$ for $\xi\approx 3$, and then it settles to values close to 0.9 for $N$-body models with initial anisotropy index near the consistency limit ($\xi\approx 7.73$). We note that the maximally rotating $N$-body models ($\alpha=1$) present lower values of $T$ with respect to their parent non-rotating OM models for $\xi\lesssim4$. In practice, weak ROI (i.e., as induced by lower values of $\xi$) typically produces nearly oblate configurations, while strong ROI (i.e., as induced by higher values of $\xi$) produces nearly prolate end states. The presence of a total non-vanishing angular momentum has no apparent influence on $T$ for $\xi\gtrsim4$, while for nearly isotropic models it induces less prolate or more oblate final states, regardless of the flattening (i.e., the minimum to maximum axial ratio $c/a$).
\begin{figure*}
    \centering
    \includegraphics[width=0.94\textwidth]{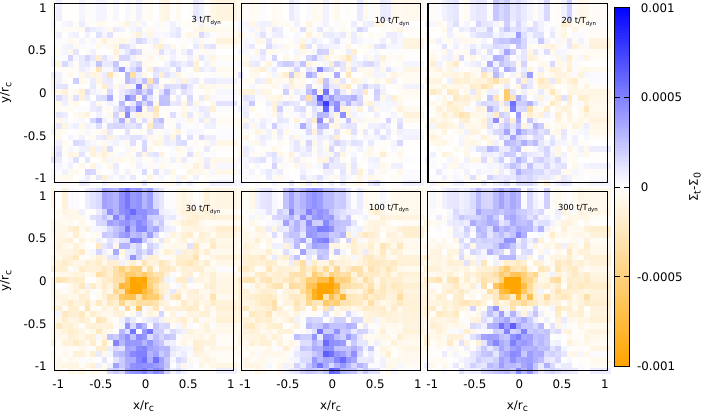}
    \caption{Time evolution of the central projected density contrast $\Sigma_t-\Sigma_0$ for the maximally anisotropic $N$-body model without net rotation ($\alpha=0$, $\xi\approx 7.73$). The spatial coordinates are expressed in units of the model's initial scale radius, $r_c$.}
    \label{fig:rhonorot}
\end{figure*}
\begin{figure*}
    \centering
    \includegraphics[width=0.92\textwidth]{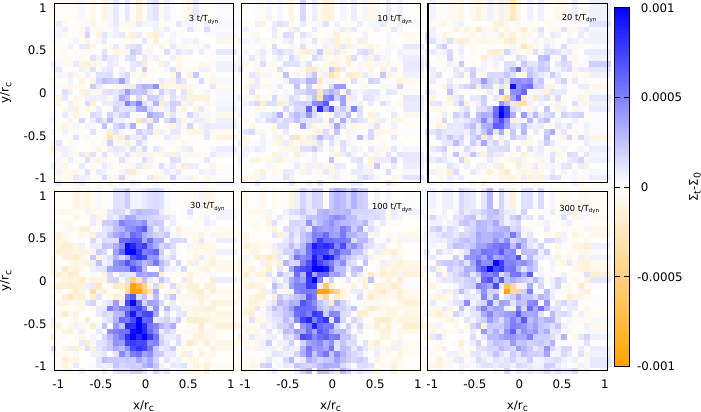}
    \caption{Same as Fig.~\ref{fig:rhonorot}, but for the maximally rotating, maximally anisotropic $N$-body model ($\alpha=1$, $\xi\approx 7.73$).}
    \label{fig:rhorot}
\end{figure*}
\subsection{Evolution of the density modes}\label{sec:modes}
\begin{figure}
    \centering
    \includegraphics[width=\columnwidth]{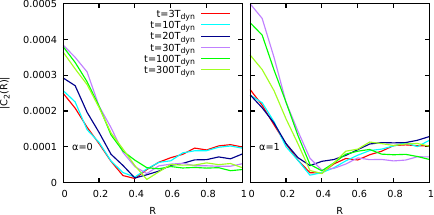}
    \caption{Time evolution of the radial profile $|C_2(R)|$ of the $n=2$ modal decomposition of $\rho$ for the non-rotating (left panel) and maximally rotating (right panel) $\xi=7.73$ models of Figs.~\ref{fig:rhonorot} and \ref{fig:rhorot}.}
    \label{fig:a2R}
\end{figure}

\begin{figure}
    \centering
    \includegraphics[width=\columnwidth]{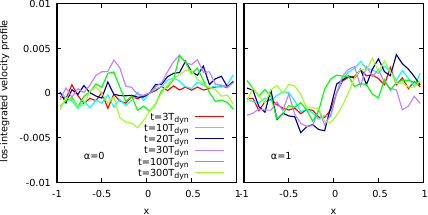}
    \caption{Time evolution of the line-of-sight integrated velocity profile for the non-rotating (left panel) and maximally rotating (right panel) $\xi=7.73$ models of Figs.~\ref{fig:rhonorot} and \ref{fig:rhorot}.}
    \label{fig:pattern}
\end{figure}

\begin{figure}
    \centering
    \includegraphics[width=\columnwidth]{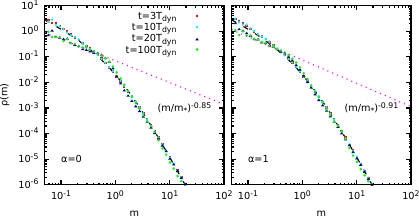}
    \caption{Time evolution of the circularised density profile for the non-rotating (left panel) and maximally rotating (right panel) $\xi=7.73$ models of Figs.~\ref{fig:rhonorot} and \ref{fig:rhorot}. In both panels the black dashed line marks the Hernquist profile (Eq. \ref{eq:hernquist}). To guide the eye, the magenta dotted lines mark the final inner density profiles.}
    \label{fig:rho2}
\end{figure}

A finer insight into the development of collective instabilities is given by the analysis of the time evolution of individual modes of the density distribution (\citealt{1991ApJ...368...66W,1994ApJ...421..481W,2025ApJ...993..150D}). For all $N$-body models, we evaluated the normalised $A_n$ modes for $n=1,2,3,4$, according to Eq.~(\ref{eq:Cn}). In Fig.~\ref{fig:modesn}, we depict such density modes (rescaled to their initial values) for the maximally anisotropic $N$-body model ($\xi=7.73$) and the nearly isotropic one ($\xi=1.60$) with $\alpha=0$ (no rotation) and $\alpha=1$ (maximal rotation). In the ROI-prone $N$-body models, the even modes $A_2$ and $A_4$ grow nearly exponentially, after a first oscillatory phase. Curiously, the presence of rotation (dashed curves) delays the onset of the exponential growth for $A_2$, while anticipating it for $A_4$. For the nearly isotropic $N$-body models, we observe instead a larger drift in all modes of the maximally rotating case. The time evolution of the $|A_1|$ mode is presented in Fig.~\ref{fig:mode1} for the same selection of $N$-body models. In the nearly isotropic case ($\xi=1.60$, right panel), the presence of rotation does not alter $|A_1|$ significantly, while for the maximally anisotropic model ($\xi=7.73$, left panel) the growth of $|A_1|$ appears to be damped by rotation, even though its net amount of rotation (i.e., the magnitude of $J_{\rm tot}^2/r_c^2$ in units of specific energy) is significantly lower with respect to the maximally rotating nearly isotropic model.\\
\indent An indicator of the response of the bulk of the system to the different types of instabilities is given by the $n=2$ mode (see, e.g., \citealt{2019MNRAS.487..711R,2021MNRAS.502.4762B}). In Fig.~\ref{fig:mode2}, we show the time evolution of $|A_2(t)|$ (rescaled to its initial value) for the same $N$-body models illustrated in Fig.~\ref{fig:ca}. At large values of $\xi$ (here 7.73, 3.77 and 2.70), consistently with the behaviour of the axial ratio $c/a$, for the maximally rotating cases ($\alpha=1$, darker blue lines), we observe a longer transient before $|A_2|$ starts growing significantly. The duration of such a phase for $N$-body models with smaller values of the rotation parameter $\alpha$ does not show any appreciable trend with it. We interpret this as an additional confirmation that rotation has a mitigating effect on the strength of the ROI. For $N$-body models with intermediate values of the anisotropy index ($\xi\approx2.36$), where a larger amount of total angular momentum at given $\alpha$ is available (see, e.g., Fig.~\ref{fig:J2xira}), the growth of the mode $|A_2|$ displays an oscillatory envelope in rotating models for $\alpha>0.1$; the corresponding modulation frequency $\omega_{\rm mod}$, not to be confused with the pattern frequency $\Omega_n$ of Eq.~(\ref{eq:omega}), does not display any significant dependence on $\alpha$. Moving towards more nearly isotropic systems, $|A_2|$ remains substantially constant over time, at least up to $10^3T_{\rm dyn}$. In Fig.~\ref{fig:gammaalphaxi}, we summarise the dependence of the growth rate of the $n=2$ mode ($\gamma_2$) on both $\alpha$ and $\xi$. In the left panel, the logarithm of $\gamma_2$ is mapped against $\alpha$ and $\xi$, qualitatively yielding a complementary picture to that of $c/a$ in Fig.~\ref{fig:caalphaxi}. From the right panel, where $\gamma_2$ is shown as a function of $\xi$, it appears that the growth rate $\gamma_2$ is an increasing function of the anisotropy as parametrised by $\xi$. Increasing $\alpha$, on average, implies a slight decrease of $\gamma_2$ for systems with large $\xi$ and an increase in the nearly isotropic cases.\\ In Appendix \ref{sec:cored} we compare these results with the stability analyses of \cite{2019MNRAS.487..711R} and \cite{2021MNRAS.502.4762B} for cored models: a cusp as shallow as the Hernquist one does not displace the stability boundary appreciably. The comparison is however restricted to $\xi\leq 2$, so that the mitigation we measure at larger anisotropy has at present no counterpart.\\ 
\indent Typically, violently unstable systems tend to suffer significant changes in the inner part of their density distribution, with notable alteration of the central density slope. Moreover, the density cusp is prone to a particle-like motion once displaced by a heavier perturber, a process dubbed ``cusp sloshing''. Following \cite{2025ApJ...993..150D} and \cite{2025arXiv251114912V}, we have evaluated in selected $N$-body models the time-dependent projected density contrast
\begin{equation}
\Delta\Sigma_t=\Sigma_t-\Sigma_0,
\end{equation}
where $\Sigma_t$ and $\Sigma_0$ are the projected densities at time $t$ and in the initial conditions\footnote{We note that other authors use the analytical projected density profile from which the initial particle positions have been sampled in lieu of $\Sigma_0$, while here we simply use its numerical realisation as for $\Sigma_t$.}; the projected densities have been computed on a $32\times32$ fixed square mesh with sides equal to $r_c$.\\
\indent In Figs.~\ref{fig:rhonorot} and \ref{fig:rhorot}, we illustrate $\Delta\Sigma_t$ for $t=3$, 10, 20, 30, 100 and $300T_{\rm dyn}$ for the non-rotating ($\alpha=0$) and maximally rotating ($\alpha=1$) $N$-body model with initial anisotropy index $\xi=7.73$. In both cases, at early times (say up to $t\approx10T_{\rm dyn}$), we observe only slight noise. At later times, the non-rotating model develops a two-lobed elongated feature with a noticeable reduction of the central density, associated with a softened inner cusp. Such a structure remains stable up to $300T_{\rm dyn}$. Conversely, in the model starting with rotating initial conditions, this non-spherical density structure is reshaped by rotation, and it evolves into a dipolar mode where an overdensity and an underdensity orbit each other (see, e.g., bottom right panel in Fig.~\ref{fig:rhorot}). This is reflected in the evolution of the radial profile $|C_2(R)|$ of the $n=2$ mode plotted in Fig.~\ref{fig:a2R}, where the $n=2$ mode of the rotating $(\alpha=1)$ case reaches a larger amplitude at $t\approx30T_{\rm dyn}$ and then declines slowly, by about 30 per cent by $t=300T_{\rm dyn}$, ending at a level comparable to that of the $\alpha=0$ model, whose amplitude remains nearly constant after $t\approx30T_{\rm dyn}$. In addition, we evaluated the corresponding integrated line-of-sight velocity profile, shown as a function of the $x$ coordinate in Fig.~\ref{fig:pattern}. A rotation pattern, antisymmetric in $x$ and persistent at all times shown, is discernible for the model with $\alpha=1$, although its amplitude is comparable to that of the fluctuations displayed by the non-rotating model. For completeness, in Fig.~\ref{fig:rho2} we show the corresponding circularised density profiles $\rho(m)$, where $m=\sqrt{(x/a)^2+(y/b)^2+(z/c)^2}$ is the ellipsoidal radius, with $a\equiv1$ in units of $r_c$ and the axial ratios $b/a$ and $c/a$ evaluated as described in Sect.~\ref{diagn}, so that $m$ reduces to $r$ for a spherical system; $m_*$ denotes the ellipsoidal radius enclosing half of the mass. We observe that, in both $\alpha=0$ and $\alpha=1$ cases, the inner logarithmic slope is shallower than that of the parent Hernquist profile. The magenta dotted lines, drawn to guide the eye through the inner part of the final profiles ($0.06\lesssim m\lesssim1$), have logarithmic slopes of $\approx0.85$ and $\approx0.91$ for $\alpha=0$ and $\alpha=1$. Over the same interval the Hernquist profile, whose local logarithmic slope $1+3m/(1+m)$ equals unity only at the centre and rises from 1.17 at $m=0.06$ to 2.5 at $m=1$, has an effective slope of $\approx1.6$. The density profiles at large $m$ match instead the radial density profile for the parent Hernquist initial condition, marked in the figures by the black dashed line. The erosion of the inner cusp is more pronounced for the non-rotating model, consistently with the evolution of the density modes.
\subsection{Evolution of the phase space distribution}
\begin{figure}
    \centering
    \includegraphics[width=\columnwidth]{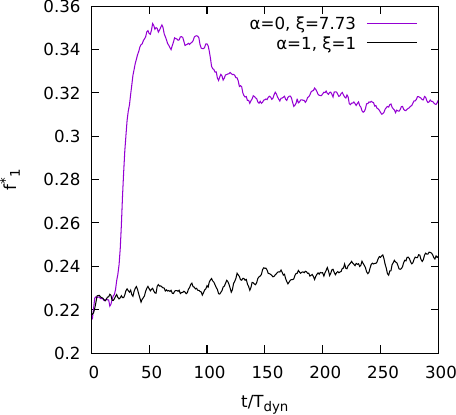}
    \caption{Time evolution of the mode $f^*_{k=1}$ of the phase space distribution for a maximally anisotropic model without rotation and a maximally rotating isotropic model.}
    \label{fig:f*1}
\end{figure}

Up to this point, we have concentrated on the response of the density profile to different degrees of rotation and anisotropy. An additional tool to probe the evolution of the $N$-body models under consideration is the analysis of the modes of their phase space distribution. For all $N$-body models in our suite, we evaluated the discrete Fourier transform of the instantaneous phase space distribution $f^*$, as given in Eq.~(\ref{eq:fourierdiscr}), where positions and velocities are expressed in cylindrical coordinates. In Fig.~\ref{fig:f*1}, we show over $300T_{\rm dyn}$ the evolution of $f^*_{k=1}$ for the two limiting cases, i.e., the non-rotating maximally anisotropic $N$-body model (purple curve) and the isotropic $N$-body model with the largest accessible angular momentum (black curve). In the ROI-prone case ($\xi=7.73$, $\alpha=0$), the $k=1$ mode of the phase space distribution displays a violent exponential growth on a time scale of about $40T_{\rm dyn}$. Conversely, in the maximally rotating isotropic case ($\xi=1$, $\alpha=1$), $f^*_{k=1}$ grows linearly. Different combinations of $\alpha$ and $\xi$ (not shown here) result in $f^*_{k=1}$ growing in a qualitatively intermediate fashion, with a behaviour leaning towards exponential or power-law growth depending on the nature of their initial conditions.

\section{Conclusions and outlook}\label{sec:conclusions}
We have investigated the stability properties of a family of Osipkov-Merritt anisotropic spherical self-gravitating systems, with net global angular momentum implemented via the ``Lynden-Bell daemon'' protocol. In contrast to previous investigations that considered only cored configurations, here we consider equilibria with a moderately cuspy central density distribution (i.e., a Hernquist sphere with an inner logarithmic density slope of unity). A comparison with previous stability analyses of the corresponding cored models is given in Appendix \ref{sec:cored}. 

The main result from the analysis of the density modes is that, in the presence of strong radial velocity anisotropy (i.e., $\xi\approx7.73$), a substantial amount of angular momentum ($\alpha\gtrsim0.5$) results in a milder radial orbit instability (similarly to what was observed by \citealt{1999MNRAS.305..859A} for anisotropic cores). Conversely, for marginally stable anisotropic equilibria (i.e., $\xi\approx 1.60$), the presence of angular momentum favours the onset of the unstable modes. 

We also characterised the morphology of the final configuration emerging from the instability phase. It appears that, for low initial values of the anisotropy index ($\xi\lesssim3$), the end states of the rotating configurations are systematically flatter than their non-rotating counterparts with the same value of $\xi$. For values of $\xi$ near the consistency limit, when the final flattening is maximal (i.e., $c/a\approx 0.3$, corresponding to an E7 system), the non-rotating models are marginally flatter than the rotating ones. Such a behaviour appears to be independent of the Lagrangian radius considered, at least up to that enclosing 90\% of the total bound mass. 

We observe that the final values of the triaxiality index $T$ have a non-monotonic behaviour as a function of $\xi$, for both rotating and non-rotating configurations, with the rotating systems having systematically lower values of $T$ at $\xi\sim 3$, thus being significantly less prolate than the non-rotating ones, for which $T\approx 1$.

We also explored the growth rates of the modes of the phase space distribution function in the limiting cases of maximally anisotropic non-rotating and isotropic maximally rotating models. We find that the modes $f^*_k$ associated with low values of $k$ grow nearly exponentially before saturating in the $\xi=7.73$ systems, and nearly linearly for $\xi\sim 1$ with maximal rotation (i.e., $\alpha\sim1$). The models with other combinations of $\xi$ and $\alpha$ display an intermediate behaviour.

Recent studies (see, e.g., \citealt{2021ApJ...912...43B,2022ApJ...926..215B,2025ApJ...993..150D}) have shown that weak cusps hosting a massive central object are prone to instability associated with the growth of non-spherical modes, eventually driving the heavy object outwards in a process often referred to as ``dynamical buoyancy''. The interplay of this collective effect with the host system's net angular momentum and/or orbital anisotropy remains to be characterised.

All cases examined here consider a radial flavour of velocity anisotropy; an equivalent analysis could be conducted for tangentially anisotropic rotating spherical equilibria (i.e., extending the investigations of \citealt{2019MNRAS.487..711R} and \citealt{2021MNRAS.502.4762B} to systems with a central density cusp). This topic will be the subject of a follow-up study, along with the introduction of an initial departure from spherical symmetry (see, e.g., \citealt{2025ApJ...986..203T,2026A&A...708A..10B}), as, in principle, no spherical model with rotation has been proved to be asymptotically stable. Moreover, it also remains to be determined how the specific details of the cuspy density profiles (i.e., the value of the inner logarithmic density slope) affect the (in)stability in the presence of rotation.
\begin{acknowledgements}
We would like to thank an anonymous Referee, Shashank Dattathri, and Giulia Congera for their careful and constructive comments. This research was supported in part by grant NSF PHY-2309135 to the Kavli Institute for Theoretical Physics (KITP). PFDC acknowledges support from the MUR PRIN2022 project ``Breakdown of ergodicity in classical and quantum many-body systems'' (BECQuMB) Grant No. 20222BHC9Z. ALV acknowledges support from a UKRI Future Leaders Fellowship (MR/S018859/1; MR/X011097/1). PFDC thanks the Royal Observatory, Edinburgh for a visit (supported by the above UKRI grant) during which this project was completed. 
\end{acknowledgements}
\bibliographystyle{aa}
\bibliography{biblio}

\begin{appendix}
\nolinenumbers
\section{Comparison with previous stability analyses of cored models}\label{sec:cored}
The rotating equilibria studied in this work are closely related to those investigated, by means of linear response and $N$-body simulations, by \cite{2019MNRAS.487..711R} and \cite{2021MNRAS.502.4762B}. In those studies the parent models are Plummer spheres, that is, systems with a central density core, while rotation is introduced exactly as done here, by applying the \cite{1960MNRAS.120..204L} daemon to a non-rotating anisotropic distribution function, with the same meaning of the parameter $\alpha$. The diagnostics also coincide: the growth rate and the pattern frequency of the unstable modes are defined there as the logarithmic derivatives of the modulus and of the argument of the $n=2$ density mode, that is, exactly as our $\gamma_n$ and $\Omega_n$ in Eqs.~(\ref{eq:gamma}) and (\ref{eq:omega}). A comparison between the two sets of results is therefore meaningful, and allows us to assess which of the features discussed in Sect.~\ref{sec:modes} are peculiar to the presence of a central density cusp.\\
\indent The two families are however labelled by different anisotropy parameters. The cored models are built on the \cite{1987MNRAS.224...13D} distribution function, whose index $q$ (not to be confused with the dimensionless energy variable of Sect.~\ref{mod1}) grows with the degree of radial bias, whereas our models are labelled by the index $\xi$ defined in Eq.~(\ref{eq:xi}). To compare them, it is convenient to express $\xi$ in terms of the anisotropy profile. Using $\sigma_t^2=2\sigma_r^2(1-\beta)$, which follows directly from Eq.~(\ref{eq:beta}), the tangential kinetic energy can be written as $K_t=4\pi\int\rho\sigma_r^2(1-\beta)r^2{\rm d}r$, so that
\begin{equation}\label{eq:xibeta}
\xi=\frac{2K_r}{K_t}=\frac{1}{1-\beta_K},\qquad
\beta_K\equiv\frac{\int\rho\,\sigma_r^2\,\beta\,r^2{\rm d}r}{\int\rho\,\sigma_r^2\,r^2{\rm d}r}.
\end{equation}
In other words, $\xi$ is set by the average of the anisotropy profile weighted by the radial kinetic energy, a quantity that can be evaluated for any model, whatever its density profile or its anisotropy parametrisation.\\
\indent For the cored models $\beta_K$ can be evaluated in closed form. Their radial velocity dispersion scales as $\sigma_r^2\propto(1+r^2)^{-1/2}$ and their density as $\rho\propto(1+r^2)^{-5/2}$, so that the weight entering Eq.~(\ref{eq:xibeta}) is proportional to $r^2(1+r^2)^{-3}$, while their anisotropy profile is $\beta(r)=(q/2)r^2/(1+r^2)$. The ratio of the two resulting integrals is exactly $1/2$, hence $\beta_K=q/4$ and
\begin{equation}\label{eq:xiq}
\xi=\frac{4}{4-q}.
\end{equation}
For our OM Hernquist models no such closed form is available, and $\beta_K$ has to be obtained numerically from the solution of the Jeans equation; we find $\beta_K=0.375$, 0.471 and 0.871 for $\xi=1.60$, 1.89 and 7.73.\\
\indent Equation (\ref{eq:xiq}) delimits the scope of the comparison. The \cite{1987MNRAS.224...13D} distribution function is consistent only for $q<2$, corresponding to $\xi<2$, whereas the most radially anisotropic OM Hernquist model considered here reaches $\xi=7.73$. The cuspy equilibria explored here therefore reach degrees of radial anisotropy that are not accessible to the cored family at all, and only our two least anisotropic models, $\xi=1.60$ and 1.89 (equivalent to $q=1.50$ and 1.88), can be set against the linear response results; the remaining ones would require $q=2.3$, 2.5, 2.9 and 3.5 for $\xi=2.36$, 2.70, 3.77 and 7.73.\\
\indent Within the range where the comparison is possible, the cusp appears to displace the stability boundary only slightly. In the absence of rotation the cored models are unstable only through the radial orbit instability, and the fitting formula given by \citeauthor{2019MNRAS.487..711R} (\citeyear{2019MNRAS.487..711R}, their Eq. 16) for the Fast-ROI vanishes at $q=1.15$, that is $\xi=1.40$. If one adopts instead the threshold on the growth rate, $\gamma_2=0.01$ in their units, below which they stopped searching for unstable modes, the boundary moves to $q=1.91$, or $\xi=1.92$. Our threshold $\xi_s\approx1.60$ falls between these two values. We conclude that an inner logarithmic density slope as shallow as the Hernquist one does not shift the onset of the instability appreciably.\\
\indent The role of the angular momentum deserves a more careful discussion, because in the cored case two distinct families of unstable modes compete with one another. \cite{2019MNRAS.487..711R} find that radially anisotropic rotating spheres support two families of unstable modes, which they name Fast-ROI and Slow-ROI: the former is the continuation of the familiar radial orbit instability into the rotating regime, while the latter rotates much more slowly and owes its existence to the net angular momentum of the system. The growth rate of the Fast-ROI obeys $\partial\gamma_2/\partial\alpha=11/100-q/109-7\alpha/25$, which is positive at low rotation and changes sign at $\alpha\approx0.34$: a modest amount of angular momentum enhances this mode, while a larger one quenches it. The Slow-ROI, by contrast, grows monotonically with $\alpha$ and overtakes the Fast-ROI at $\alpha\approx0.5$, attaining at $\alpha=1$ a growth rate about 75 per cent larger than the maximum ever reached by the Fast-ROI. In the cored case, therefore, rotation does not stabilise the system: it substitutes one instability for another, and the dominant mode at high rotation is the Slow-ROI.\\
\indent This behaviour is consistent with what we observe for our least anisotropic models, for which increasing $\alpha$ anticipates the departure from spherical symmetry. The slight mitigation that we find instead at $\xi\gtrsim4$ has no counterpart in the cored analysis, simply because those anisotropies lie beyond the consistency limit of the cored family. Whether the quenching of the Fast-ROI is accompanied there, as it is in the cored case, by the emergence of a Slow-ROI taking its place, or whether the cusp alters the competition between the two, cannot be decided on the basis of the present simulations. We note that an extension of the linear response charting to cuspy equilibria was already indicated as a desirable development by \citeauthor{2019MNRAS.487..711R} (\citeyear{2019MNRAS.487..711R}).\\
\indent Finally, the behaviour outlined above admits a qualitative interpretation in terms of the precession picture invoked in Sect.~\ref{intro}. The relevant quantity is the distribution of the tumbling frequencies $\Omega_{\rm ILR}=\Omega_\varphi-\Omega_r/2$ of the orbits, where $\Omega_\varphi$ and $\Omega_r$ are the azimuthal and radial orbital frequencies, an instability being favoured when such a distribution is sufficiently narrow (\citealt{1979MNRAS.187..101L}). In a non-rotating model the distribution is symmetric, since the phase space distribution function does not depend on the sign of $J_z$. The application of the \cite{1960MNRAS.120..204L} daemon multiplies its retrograde branch by $1-\alpha$ and its prograde branch by $1+\alpha$, so that a small amount of rotation suppresses the counter-precessing orbits and narrows the distribution, thereby favouring the onset of the instability. As $\alpha$ increases further, however, the growing mode acquires a finite pattern speed, and the resonance condition is met by an ever smaller fraction of the remaining orbits: the trapping of the radial orbits onto the aspherical perturbation is progressively detuned, and the Fast-ROI is weakened. This argument accounts for the behaviour of that mode alone, and says nothing about the Slow-ROI, whose growth relies instead on the rotational kinetic energy of the system.
\end{appendix}
\end{document}